\documentclass[1p]{elsarticle}

\usepackage[english]{babel}
\usepackage{amsfonts}
\usepackage{amssymb}
\usepackage{amsmath}
\usepackage{amsthm}
\usepackage{latexsym}
\usepackage{graphicx}
\usepackage{natbib}

\def\LL{{\cal L}}
\def\OO{{\cal O}}

\journal{Physics Letters B}

\begin{document}

\begin{frontmatter}

\title{Perturbation of the metric around a spherical body from a nonminimal coupling between matter and curvature}

\author{Nuno Castel-Branco}
\ead{nuno.castel-branco@tecnico.ulisboa.pt}
\address{Instituto Superior T\'ecnico, Universidade de Lisboa,Av. Rovisco Pais 1, 1049-001 Lisboa, Portugal}

\author{Jorge P\'aramos}
\ead{jorge.paramos@fc.up.pt}
\address{Centro de F\'isica do Porto and Departamento de F\'isica e Astronomia, Faculdade de Ci\^encias, Universidade do Porto, Rua do Campo Alegre 687, 4169-007 , Porto, Portugal}

\author{Riccardo March}
\ead{r.march@iac.cnr.it}
\address{Istituto per le Applicazioni del Calcolo, CNR, Via dei Taurini 19, 00185 Roma, Italy, and INFN - Laboratori Nazionali di Frascati (LNF), via E. Fermi 40, Frascati 00044 Roma, Italy.}

\date{\today}

\begin{abstract}
In this work, the effects of a nonminimally coupled model of gravity on a perturbed Minkowski metric are presented. The action functional of the model involves two functions $f^1(R)$ and $f^2(R)$ of the Ricci scalar curvature $R$.
Based upon a Taylor expansion around $R=0$ for both functions $f^1(R)$ and $f^2(R)$, we find that the metric around a spherical object is a perturbation of the weak-field Schwarzschild metric: the time perturbation is shown to be a Newtonian plus Yukawa term, which can be constrained using the available experimental results. We conclude that the Starobinsky model for inflation complemented with a generalized preheating mechanism is not experimentally constrained by observations. The geodetic precession effects of the model are also shown to be of no relevance for the constraints.
\end{abstract}


\end{frontmatter}

\section{Introduction}

Modern physics uses the concepts of dark matter and dark energy to advance an explanation for the astrophysical problem of the flattening of galactic rotation curves and the cosmological problem of the accelerated expansion of the universe, respectively. Dark energy, which is supposed to account for 74$\%$ of all the matter of the universe, has many theories on its basis, as the so-called "quintessence" models \cite{quintessence1,quintessence2,quintessence3} and the existence of scalar fields that account for both dark matter and dark energy \cite{scalarfield}.

More recent approaches start from the idea of the incompleteness of the fundamental laws of General Relativity (GR), involving, for example, corrections to the Einstein-Hilbert action. Such theories involve a nonlinear correction to the geometry part of the action, being thus called $f(R)$ theories. In the last decade, work on $f(R)$ theories has been very profitable, as thoroughly discussed in Ref. \cite{felice}. These can be extended to also include a nonminimum coupling (NMC) between the scalar curvature and the matter Lagrangian density.

Indeed, these NMC theories have many interesting features, as can be seen by several studies, such as the impact on stellar observables \cite{stelobserv}, the energy conditions \cite{energcondit}, the equivalence with multi-scalar-tensor theories \cite{multiscalar}, the possibility to account for galactic \cite{drkmattgal} and cluster \cite{drkmattclus} dark matter, cosmological perturbations \cite{cosmpertur}, a mechanism for mimicking a Cosmological Constant at astrophysical scales \cite{mimlambda}, post-inflationary reheating \cite{reheating} or the current accelerated expansion of the universe \cite{curraccel,Friedmann}, the dynamical impact of the choice of the Lagrangian density of matter \cite{dynimpac1,dynimpac2}, gravitational collapse \cite{gravcollapse}, its Newtonian limit \cite{newtlimit}, the existence of closed timelike curves \cite{closedtimecurve} and, the most recent one, a determination of solar system constraints to a cosmological NMC \cite{solar}.

For other NMC gravity theories and their potential applications, see {\it e.g.} \cite{puetz1,puetz2,puetz3,obuk,iorio}.

One of the first motivations that brought $f(R)$ theories into the physicists daily work was the Starobinsky inflation model, where $f(R)=R+R^2/(6m^2)$ was considered \cite{starobinsky,reheating}, with WMAP normalization of the CMB temperature anisotropies indicating that $m\sim 3 \times 10^{-6} M_P$, where $M_P$ is the Planck mass \cite{planckmass}.

Without directly mentioning the Starobinsky inflation, Ref. \cite{naf} considers a quadratic $f(R)$ function and develops an expansion in powers of $ (1/c)$ of an asymptotically flat Minkowski metric, showing the presence of a Yukawa correction to the $tt$ component of the latter \cite{naf}. Following the equivalence between scalar-tensor and $f(R)$ theories \cite{analogy1,analogy2,analogy3}, this can be interpreted as due to the additional gravitational contribution of the massive degree of freedom embodied in a non-linear $f(R)$ function.

In this work we follow a similar procedure of Ref. \cite{naf} where we instead consider a NMC model. We consider that the additional degree of freedom arising from a non-trivial $f(R)$ function is sufficiently massive so that its effects are not extremely long-ranged, and as such we can neglect the background cosmological setting --- this point (which shall be developed in the following) shows that this work is complementary to the recent study on the compatibility between cosmological and solar system dynamics of a NMC model \cite{solar}.

In section II such a model is presented and in section III the solution of the linearized field equations is computed. We obtain the solutions for the perturbative potentials $\Psi(r)$ and $\Phi(r)$ of the metric, which contain a form factor specific of the Yukawa potential that is addressed in section IV.
The $tt$ component of the metric yields the modified gravitational potential, which includes a Newtonian plus a Yukawa contribution. The comparison of these results with available experimental constrains is presented in section V. This section also addresses the radial potential through the constraints obtained to the geodetic precession values. Finally, conclusions are drawn.

\section{The model}

The action functional of gravity for the NMC case is of the form \cite{BBHL}
\begin{equation}\label{model}
S=\int \left[\dfrac{1}{2} f^1(R)+\left[1+f^2(R)\right] \LL_m\right]
\sqrt{-g} d^4x,
\end{equation}
\noindent where $f^i(R) \left(i=1,2\right)$ are functions of the Ricci scalar curvature $R$, $\LL_m$ is the Lagrangian density of matter and $g$ is the metric determinant. The standard Einstein-Hilbert action is recovered by taking
\begin{equation}
\begin{matrix}
f^1(R)=2\kappa(R-2\Lambda),	&&	f^2(R)=0,
\end{matrix}
\end{equation}
\noindent where $\kappa=c^4/16\pi G$, $G$ is Newton's gravitational constant and $\Lambda$ is the Cosmological Constant.

The variation of the action functional with respect to the metric $g_{\mu\nu}$ yields the field equations
\begin{equation} \label{field equations}
\left(f^1_R + 2 f^2_R \LL_m \right) R_{\mu \nu} -
\dfrac{1}{2} f^1 g_{\mu \nu} =
\left(1+ f^2 \right) T_{\mu \nu} +\left( \square_{\mu \nu} - g_{\mu \nu} \square \right)
\left( f^1_R + 2 f^2_R \LL_m \right),
\end{equation}
\noindent where $f^i_R\equiv df^i \slash dR$ and $\square_{\mu\nu}\equiv \nabla_\mu \nabla_\nu$.

In the following we assume that matter behaves as dust, $i.e.$ a perfect fluid with negligible pressure and an energy-momentum tensor described by
\begin{equation}
T_{\mu\nu} = \rho c^2 u_\mu u_\nu~~~~,~~~~  u_\mu u^\mu =-1,
\end{equation}
\noindent where $\rho$ is the matter density and $u_\mu$ is the four-velocity vector. The trace of the energy-momentum tensor
is $T = -\rho c^2$. We use $\LL_m = -\rho c^2$ for the Lagrangian density of matter (see Ref. \cite{dynimpac1} for a discussion).

We consider a spherically symmetric body with a static radial mass density $\rho = \rho(r)$ and we assume that the function
$\rho(r)$ and its first derivative are continuous across the surface of the body,
\begin{equation} \label{conditions for fe}
\rho(R_S) = 0 \quad \text{and} \quad \dfrac{d\rho}{dr}(R_S) = 0,
\end{equation}
\noindent where $R_S$ denotes the radius of the spherical body. These conditions will play a crucial role in the following sections, when integrals that have $R_S$ as an integration limit will appear.

The metric used is one that describes the spacetime around a spherical star like the Sun and it is given by the following perturbation of the Minkowski metric, in spherical coordinates:
\begin{equation} \label{metric}
ds^2=-\left[1+2\Psi\left(r\right)\right]c^2dt^2+\left[1+2\Phi(r)\right] dr^2 + r^2 d\Omega^2,
\end{equation}
\noindent where $\Psi$ and $\Phi$ are perturbing functions such that $\left|\Psi(r)\right| \ll 1$ and $\left|\Phi(r)\right| \ll 1$.

For the purpose of the present paper the functions $\Psi$ and $\Phi$ will be computed at order $\OO(1/c^2)$.

We assume that the functions $f^i(R)$ admit the following Taylor expansions around $R=0$, which coincide with the forms used in Ref. \cite{reheating}:
\begin{equation} \label{f(R) equations}
f^1(R)=2\kappa \left(R+\dfrac{R^2}{6m^2}\right) + \OO(R^3), \qquad\qquad
f^2(R)=2\xi \dfrac{R}{m^2} + \OO(R^2),
\end{equation}
\noindent where $m$ is a characteristic mass scale and $\xi$ a dimensionless parameter specific of the NMC, indicating the relative strength of the latter with respect to the quadratic term in $f^1(R)$.

Notice also that the Cosmological Constant is dropped, consistent with the assumption that the metric is asymptotically flat --- {\it i.e.} no cosmological background with a time-dependent, non-vanishing curvature $R_0 \neq 0$ is assumed, contrary to what was considered in Ref. \cite{solar}. In that study, a set of viability criteria for the form of $f^2(R)$ was developed based upon the compatibility of the large scale effects ({\it i.e.} description of dark energy) and allowed solar system impact: it is worth mentioning that the validity of such criteria required a very light additional degree of freedom, $m_0 r \ll 1$, with mass given by
\begin{equation}\label{mass-formula}
m_0^2 =  \frac{1}{3}\bigg[\frac{f^1_{R0} -  f^2_{R0}\LL_m}{f^1_{RR0} + 2 f^2_{RR0}\LL_m}
- R_0 -  \frac{3\square\left( f^1_{RR0} - 2 f^2_{RR0}\rho^{\rm cos}\right)
- 6\rho\square f^2_{RR0}}{f^1_{RR0} + 2 f^2_{RR0}\LL_m} \bigg],
\end{equation}
\noindent where the subscript $_0$ indicates that the quantities are evaluated at their background cosmological value $R=R_0$ ({\it e.g.} $f^i_{R0} \equiv f^i_R (R_0)$) and $\rho^{\rm cos}$ is the corresponding background cosmological density.

\section{Solution of linearized modified field equations}

\subsection{Solution for the curvature $R$}

The trace of the field equations (\ref{field equations}) is
\begin{equation} \label{trace of field equations}
\left(f^1_R + 2 f^2_R \LL_m \right) R - 2 f^1 =-3 \square
\left(f^1_R + 2 f^2_R \LL_m \right)
+ \left( 1+ f^2 \right) T.
\end{equation}
\noindent After expanding the trace with the respective expressions, the equation is linearized: this is done by neglecting terms of order $\OO( 1 \slash c^3)$ or smaller. It yields the following equation,
\begin{equation} \label{curvature equation}
\nabla^2 R - m^2 R = - \dfrac{8\pi G}{c^2} m^2
\left[ \rho -6\left(\dfrac{2\xi}{m^2}\right) \nabla^2 \rho \right],
\end{equation}
\noindent which, by the variable substitution $u = r R$, enables the more straightforward expression
\begin{equation}
\dfrac{d^2 u(r)}{dr^2} - m^2 u(r) = s(r),
\label{curvature equation, variable substitution}
\end{equation}
\noindent where
\begin{equation}
s(r) = - \dfrac{8\pi G}{c^2} m^2 r \left[\rho - 6 \left(\dfrac{2\xi}{m^2}\right) \nabla^2 \rho \right],
\end{equation}
\noindent is the source function.

The boundary conditions on $u(r)$ are
\begin{itemize}
\item $u(0) = 0$, so that the curvature $R$ may have a finite value everywhere;
\item $\lim_{r \to \infty} u(r)\slash r = 0$, so that the curvature vanishes asymptotically as one recovers the Minkowski metric.
\end{itemize}
The Green function of Eq. \eqref{curvature equation, variable substitution} solves the following equation in the sense of distributions
\begin{equation}
\dfrac{d^2 G(r,r')}{dr^2} - m^2 G(r,r') = \delta (r - r'),
\end{equation}
\noindent where $\delta(r-r')$ is the Dirac delta distribution. The Green function $G(r,r')$ is used to determine the solution of the curvature equation in the form \eqref{curvature equation, variable substitution} by means of the integral $u(r) = \int^{R_S}_0 G(r,r') s(r') dr'$.
Due to the different boundary conditions, the curvature is written as a twofold solution:
\begin{equation} \label{curvature solution}
R(r) = \left\{
  \begin{array}{l l}
     R^\uparrow(r) & \quad \text{if  } r>R_S\\
     \quad \\
     R^\downarrow(r) & \quad \text{if  } 0\leq r\leq R_S   \end{array} \right. ,
\end{equation}
\noindent where $R^\downarrow(r)$ is the curvature inside the star,
\begin{equation}
R^\downarrow (r) = -\dfrac{4 \pi G}{c^2 m} \left[ \dfrac{e^{-mr}}{r} I_1(r)
+ \dfrac{2 \sinh(mr)}{r} I_2(r) \right],
\end{equation}
\noindent the quantities $I_1(r)$ and $I_2(r)$ have been computed by using the properties (\ref{conditions for fe})
of the mass density $\rho(r)$ and are given by
\begin{eqnarray}
I_1(r)&=& 2\left(12\xi -1\right)m^2\int^{r}_{0}\sinh(mr')r'\rho(r')dr'  -24\xi \cosh(mr)mr\rho(r),\\ \nonumber
I_2(r)&=& (12\xi-1) m^2  \int^{R_S}_r e^{-mr'}r'\rho(r')  dr' -12\xi e^{-mr} mr \rho(r)  ,
\end{eqnarray}
\noindent and $R^\uparrow(r)$ is the curvature outside the star, given by
\begin{equation} \label{curvature solution, outside star}
R^\uparrow (r) = \dfrac{2 G M_S}{c^2 r} m^2 \left( 1 - 12\xi \right) A(m,R_S) e^{-mr},
\end{equation}
\noindent with $M_S$ the mass of the spherical body and $A(m,R_S)$ a form factor defined as
\begin{equation}
A(m,R_S) = \dfrac{4\pi}{m M_S} \int^{R_S}_0 \sinh(mr) r \rho(r) dr,
\end{equation}
\noindent which will be discussed in a subsequent section.

The expression (\ref{curvature solution, outside star}) vanishes as $r\rightarrow\infty$ and it
is considered to be valid only at Solar System scales, since spacetime should assume a de Sitter metric with curvature $R_0 \neq 0$
at cosmological scales. Note also that in the limit $m \rightarrow 0$ we have $R^\uparrow(r) \rightarrow 0$ for any $r > R_S$.

\subsection{Solution for $\Psi$}

The $tt$ component of the Ricci tensor at order $\OO(1/c^2)$ is given by
\begin{equation} \label{elements, tt component}
R_{tt} = \dfrac{2\Psi'}{r}+\Psi'' + \OO\left(\dfrac{1}{c^3}\right) = \nabla^2\Psi + \OO\left(\dfrac{1}{c^3}\right).
\end{equation}
\noindent  Then, neglecting all terms smaller than $\OO(1/c^2)$ in the $tt$ component of the field equations \eqref{field equations},
and using the curvature equation \eqref{curvature equation}, we get the equation ruling $\Psi$:
\begin{equation} \label{psi equation}
\nabla^2\Psi = \dfrac{c^2}{3k}\rho-
\dfrac{R}{6}.
\end{equation}
This is solved outside the spherical body, $r \geq R_S$, by decomposing it into a sum,
\begin{equation} \label{psi equation, decomposition}
\nabla^2 \Psi = \nabla^2 \Psi_0 + \nabla^2 \Psi_1,
\end{equation}
\noindent where
\begin{equation} \label{psi 1 and psi 2}
\nabla^2 \Psi_0 = \dfrac{c^2}{3k} \rho~~~~,~~~~\nabla^2 \Psi_1 = -\dfrac{R}{6}.
\end{equation}
The first equation is easily solved using the divergence theorem, which gives
\begin{equation} \label{psi 0}
\Psi_0 (r) = - \dfrac{4 }{3 } \dfrac{GM_S}{c^2r}.
\end{equation}
The second one is more cumbersome: a tedious integration eventually shows that
\begin{equation} \label{psi 1}
\Psi_1 (r) = \dfrac{G M_S}{3 c^2 r} \left[ 1 - \left( 1 - 12 \xi \right) A(m,R_S) e^{-mr} \right],
\end{equation}
\noindent so that for $r \geq R_S$ the solution is
\begin{equation} \label{psi solution}
\Psi(r) = - \dfrac{G M_S}{c^2 r} \left[1 + \left(\dfrac{1}{3} - 4\xi \right)
A(m,R_S) e^{-mr}\right].
\end{equation}

\subsection{Solution for $\Phi$}

By the same token, being the mass distribution static, we insert the expressions
\begin{equation} \label{radial components}
R_{rr} = \dfrac{2}{r}\Phi' - \Psi'' + \OO\left(\dfrac{1}{c^3}\right)~~~~,~~~~
T_{rr} = 0,
\end{equation}
\noindent into the $rr$ component of the field equations, along with the $f^i(R)$ expressions from \eqref{f(R) equations}, obtaining
\begin{equation} \label{phi equation}
2\Phi' - r \Psi'' - \dfrac{r R}{2} + \dfrac{2 R'}{3 m^2} -
\dfrac{4 \xi c^2}{m^2 k} \rho' = 0.
\end{equation}
\noindent This equation is easily integrated outside the spherical body, $r \geq R_S$, leading to
\begin{equation} \label{phi solution}
\Phi(r)=\dfrac{G M_S}{c^2 r} \left[ 1 - \left(\dfrac{1}{3}-4\xi\right) A(m,R_S) e^{-mr} (1 +mr) \right].
\end{equation}
\noindent In the GR limit, $\xi = 0$ and $m \to \infty$, the exponential term in both $\Psi$ and $\Phi$ vanishes and we recover the weak-field approximation of the Schwarzschild metric, as expected.

\section{Discussion of Yukawa potential}

As it has been shown, from the $tt$ component of the metric we identify a Newtonian potential plus a Yukawa perturbation:
\begin{equation} \label{yukawa potential}
U(r) = - \dfrac{G M_S}{r} \left( 1 + \alpha A(m,R_S) e^{-r/\lambda} \right),
\end{equation}
\noindent defining the characteristic length $\lambda = 1/m$ and the strength of the Yukawa addition
\begin{equation} \label{alpha as the yukawa potential strength}
\alpha = \dfrac{1}{3} - 4\xi ,
\end{equation}
\noindent so that, if $\xi = 0$, we get the Yukawa strength for pure $f(R)$ theories, $\alpha = 1/3$ \cite{naf}; also, notice that a positive NMC (as assumed in Refs. \cite{stelobserv,gravcollapse}) yields $\alpha \leq 1/3$. Strikingly, a NMC with $\xi = 1/12$ cancels the Yukawa contribution.

\subsection{The form factor $A(m,R_S)$}

As defined before, the form factor is
\begin{equation} \label{form factor A(m,R_S)}
A(m,R_S) = \dfrac{4\pi}{m M_S} \int_0^{R_S} \sinh(mr) r\rho(r)dr.
\end{equation}
This dimensionless form factor was found by integrating the field equations of NMC gravity but it is not specific of the NMC gravity nor of $f(R)$ theories of this kind, but of any Yukawa model \cite{fischbach}, as it arises from the integral
\begin{equation} \label{yukawa potential, integral}
U_Y \left( \vec{x} \right) = - G \alpha \int_{B_{R_S}} \dfrac{\exp\left[-m \left| \vec{x}-\vec{x}'\right| \right]}{\left| \vec{x}-\vec{x}'\right|} \rho \left(\vec{x}'\right) d^3x',
\end{equation}
\noindent where $B_{R_S}$ is a sphere with radius $R_S$ and center at the origin. In the case of a spherically symmetric distribution of mass $\rho(r)$, evaluation of the integral \eqref{yukawa potential, integral} in spherical coordinates yields the Yukawa contribution to $U(r)$ in Eq. \eqref{yukawa potential}, so that
$G\alpha$ can be interpreted as the strength of the Yukawa potential generated by a point source.

The form factor can be evaluated in several ways, according to the function of mass density $\rho(r)$. Taking the limit of a point source, $r \to 0$ allows us to expand around $m r \ll 1$, so that $\sinh \left( m r \right) \approx m r [1 + (mr)^2/6]$ and
\begin{equation}
A(m,R_S) \approx \dfrac{4\pi}{M_S} \int^{R_S}_0 \left[1 + \dfrac{(mr)^2 }{6}\right] r^2 \rho(r) dr
=  1 + \dfrac{2m^2\pi}{ 3M_S} \int^{R_S}_0 r^4 \rho(r) dr \sim 1.
\end{equation}
\noindent This can be verified explicitly by making all computations with a test mass density (such as a uniform profile) and, in the end, taking the limit $R_S \to 0$. Indeed, taking
\begin{equation}
\rho_0 = \dfrac{3M_S}{4\pi R_S^3}
\end{equation}
\noindent we obtain
\begin{equation} \label{form factor A(m,R_S) constant}
A(m,R_S) =3 \dfrac{mR_S \cosh (mR_S) - \sinh (mR_S)}{(m R_S)^3} ,
\end{equation}
\noindent which admits the limiting cases
\begin{eqnarray} \label{form factor A(m,R_S) constant limiting}
A(m,R_S) &\approx & 1 + \dfrac{(mR_S)^2 }{ 10} \sim 1 ~~,~~mR_S \ll 1, \\ \nonumber A(m,R_S) &\approx & \dfrac{3}{2}\dfrac{e^{mR_S}}{(mR_S)^2} ~~,~~mR_S \gg 1.
 \end{eqnarray}
If the central body is the Sun (with radius $R_\odot$), we may instead consider the more accurate density profile \cite{NASAprofile} (which obeys condition $\rho(R_\odot)=0$, while $(d\rho\slash dr)(R_\odot) \simeq 0$),
\begin{equation}\label{NASA density profile}
\rho(r) = \rho_0 \bigg[  1 - 5.74 \left(\dfrac{r}{R_\odot}\right)+ 11.9\left(\dfrac{r}{R_\odot}\right)^2 - 10.5\left(\dfrac{r}{R_\odot}\right)^3 + 3.34\left(\dfrac{r}{R_\odot}\right)^4\bigg] ,
 \end{equation}
\noindent obtaining
\begin{eqnarray} \label{form factor A(m,R_S) NASA}
A(m,R_\odot) &=& x^{-7} [4.6 \times 10^4 x + 2.1 \times 10^3 x^3 +  (2.7 \times 10^4 + 131 x^2) x \cosh x - \\ \nonumber &&  (7.3 \times 10^4 + 3.6 \times 10^3x^2 - 14.6 x^4) \sinh x ],
\end{eqnarray}
\noindent (with $x= mR_\odot$, for brevity), with the limiting cases
\begin{eqnarray} \label{form factor A(m,R_S) NASA limiting}
A(m,R_\odot) &\approx & 1 + 6 \times 10^{-2} (m R_\odot)^2 \sim 1 ~~,~~mR_\odot \ll 1, \nonumber \\ A(m,R_\odot) &\approx & \ 7.3 \dfrac{e^{mR_\odot}}{(mR_\odot)^3 }~~,~~mR_\odot \gg 1.
 \end{eqnarray}

\begin{figure}[ht]
\centering
\includegraphics[width=\textwidth]{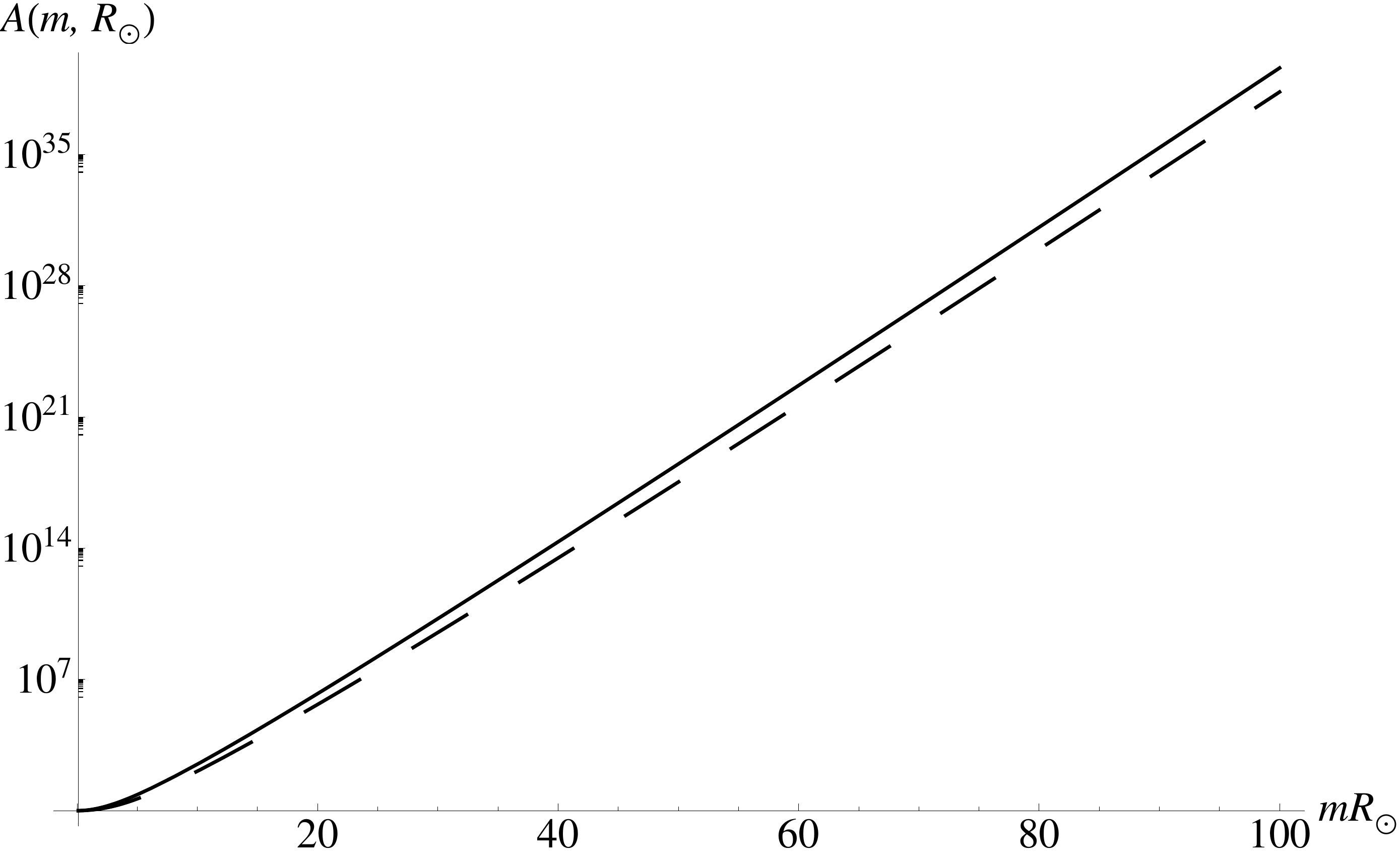}
\caption{Form factor $A(m,R_\odot)$ for a constant density (full) and fourth-order density profile for the Sun, Eq. (\ref{NASA density profile}) (dashed).}
\label{fig:formfactors}
\end{figure}

Both forms for $A(m,R_\odot)$ are plotted in Fig. \ref{fig:formfactors}, showing that it grows with $m$. Although, for values of the lengthscale $\lambda \ll R_\odot$, this effectively boosts the form factor, the contribution from the Yukawa term in Eq. (\ref{yukawa potential}) is nevertheless suppressed by the factor $\exp(-r/\lambda)$.

\subsection{PPN Parameters}

Similarly to the present work, the Parameterized Post-Newtonian formalism posits an expansion of the metric elements and other quantities (energy-momentum tensor, equations of motion, {\it etc}.) in powers of $1/c^2$ \cite{Will}; the eponymous PPN metric reads, for a spherical central body,
\begin{equation} \label{metricPPN} ds^2 = - \left[ 1 - 2\dfrac{GM_S }{c^2 r} + 2 \beta \left(\dfrac{ GM_S }{c^2 r}\right)^2 \right]c^2 dt^2 + \left( 1 + 2 \gamma \dfrac{GM_S }{c^2 r} \right)~\left( dr^2 + r^2 d\Omega^2 \right) ~~,\end{equation}
\noindent where $\beta$ and $\gamma$ are two PPN parameters, which measure the amount of non-linearity affecting the superposition law for gravity and the spatial curvature per unit mass, respectively; GR is signalled by $\beta = \gamma =1$. Other PPN parameters also appear in a more evolved version of the metric above, signalling violation of momentum conservation, existence of a privileged reference frame, amongst others deviations from GR.

Clearly, such a formalism is incompatible with the presence of a Yukawa term in the gravitational potential, since the latter cannot be expanded in powers of $1/r$; furthermore, the discussion after Eq. (\ref{curvature solution, outside star}) highlights that, in the limit $m\to 0$, we must consider the background cosmological curvature and cannot assume the asymptotically flat {\it Ansatz} (\ref{metric}) for the metric: this was performed in Ref. \cite{solar}, as already mentioned.

Nevertheless, we can consider what happens if the condition $mr \ll 1$ is valid throughout the region of interest ({\it e.g.} the Solar System), for consistency: in this case, the metric (\ref{metric}) with the solutions (\ref{psi solution}) and (\ref{phi solution}) is well approximated by
\begin{equation} \label{metriclight}
ds^2 = -\left[1 - \dfrac{2G M_S}{c^2 r} \left(\dfrac{4}{3} - 4\xi \right) \right]c^2dt^2+ \left[1+\dfrac{2G M_S}{c^2 r} \left( \dfrac{2}{3}+ 4\xi \right)\right] dr^2 + r^2 d\Omega^2,
\end{equation}
\noindent which, upon comparison with Eq. (\ref{metricPPN}) (or, mathematically, the adequate constant rescaling of both time and radial coordinates), yields
\begin{equation}\label{gamma}
\gamma= \dfrac{1 }{ 2} \dfrac{1 + 6\xi }{ 1 - 3\xi}.
\end{equation}
In the absence of a NMC, $\xi = 0 $, this yields $\gamma = 1/2$, a strong departure from GR that is disallowed by current experimental bounds, $\gamma = 1 + (2.1 \pm 2.3) \times 10^{-5}$ \cite{status}. This apparent disagreement between $f(R)$ theories and observations was noted early on (as discussed {\it e.g.} in Refs. \cite{PPNfR1,PPNfR2,CSE}), and can be avoided if the additional degree of freedom arising from a non-linear $f(R)$ function is massive enough.

The expression above appears to show that a NMC allows $f(R)$ theories to remain compatible with observations, as long as $\xi=1/12$ --- which is just a restatement of the previously obtained result. Again, the path towards obtaining the $\gamma$ PPN parameter depicted above is presented for illustration only, as it relies on an approximation of a Yukawa perturbation and disregards the fact that, in the limit $mr \ll 1$, the background cosmological dynamics cannot be neglected. As such, no conclusions can be drawn from comparison with the experimental bound on $\gamma$ mentioned above.

\section{Experimental constraints to NMC gravity parameters}

\begin{figure}[ht]
\centering
\includegraphics[width=\textwidth]{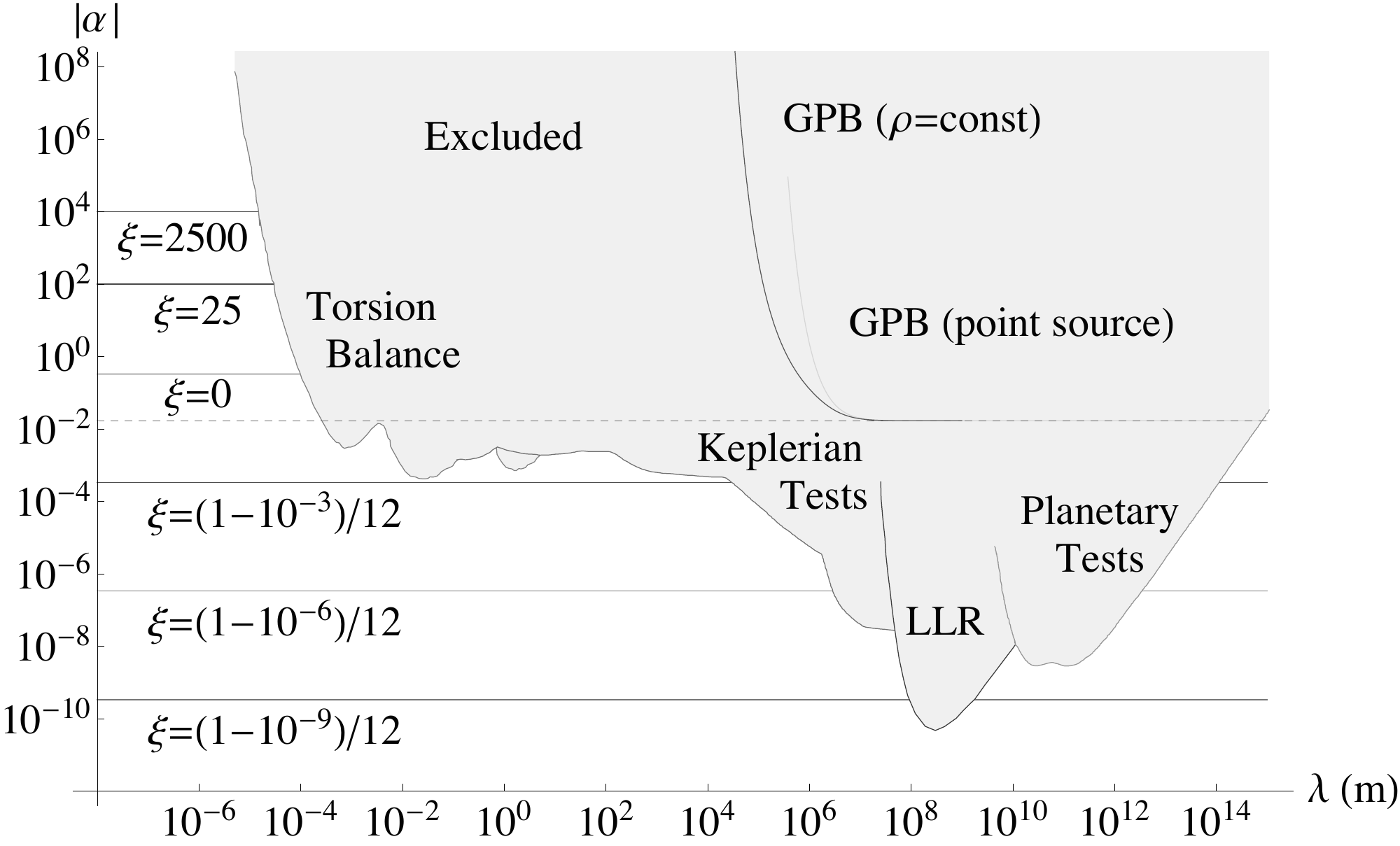}
\caption{Yukawa exclusion plot for $\alpha$ and $\lambda$. Adapted from Refs. \cite{adelberg1,salumbides}.}
\label{fig:exclusionplot}
\end{figure}

The Yukawa potential \eqref{yukawa potential} is not new in physics as an alternative way to account for deviations from Newtonian gravity or other forces of nature \cite{adelberg1,fischbach,survive,status}. Fig. \ref{fig:exclusionplot} shows the exclusion plot for the phase space $(\lambda,\alpha)$, which may be used to constraint the phase space of the model (\ref{model}) under scrutiny.

In doing so, we recall that the results obtained in Eqs. (\ref{psi solution},\ref{phi solution},\ref{yukawa potential}) are not exact, but only accurate to order $\OO(c^{-2})$, and are based upon the assumption of a perturbation to a Minkowski metric: a future analysis should expand this framework to also include terms $\OO(c^{-4})$, as well as establish some matching criteria between the static, spherically symmetric spacetime here considered and the evolving background spacetime \cite{matching} (see also Ref. \cite{darkmatter}). Indeed, Ref. \cite{Clifton} has found that in $f(R)$ theories, $\OO(c^{-4})$ terms can arise that are not exponentially suppressed, and as such may play a role at large distances, particularly if the $\OO(c^{-2})$ Yukawa interaction here obtained is short-ranged.

From Fig. \ref{fig:exclusionplot} (as discussed after \eqref{alpha as the yukawa potential strength}), one sees a NMC with $\xi = 1/12$ cancels this contribution, as shown by the values of $|\alpha| \to 0$ overlaid on the exclusion plot. Also, notice that large values of $\xi$ lead to a large, negative strength $\alpha \sim -4\xi$ (the cases $\xi =25$ and $\xi = 2500$ are shown).

\begin{figure}[ht]
\centering
\includegraphics[width=\textwidth]{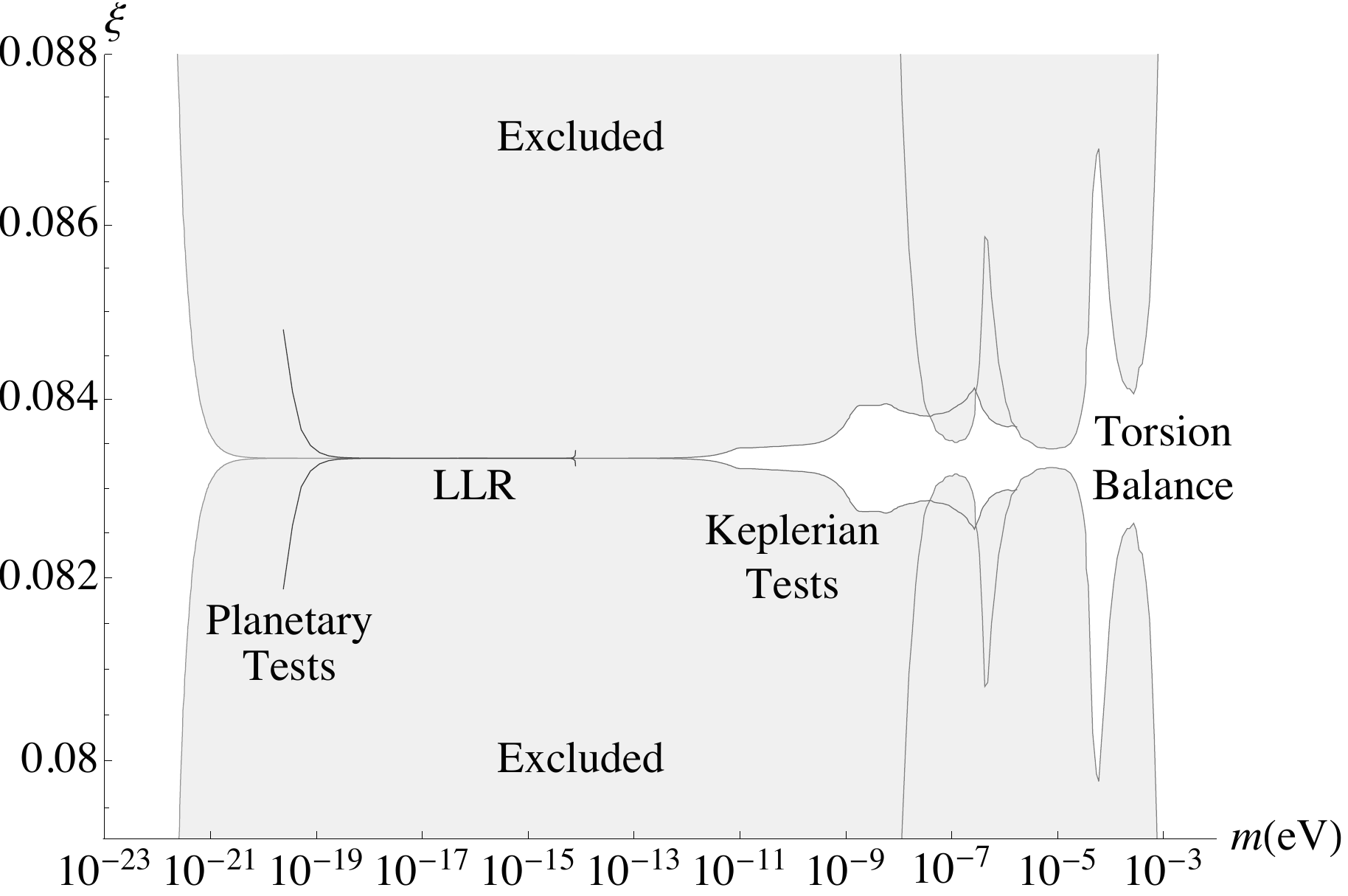}
\caption{Exclusion plot for the dimensionless relative strength $\xi$ and characteristic mass scale $m$.}
\label{fig:exclusionplotnmccoord}
\end{figure}

We may transform into the phase space $(m,\xi)$ of the model under scrutiny, \eqref{f(R) equations}, using
\begin{equation} \label{transformations, nmc graphic}
m = \frac{1}{\lambda}, \qquad \qquad
\xi = \frac{1}{12}-\frac{\alpha}{4},
\end{equation}
\noindent to get the exclusion plot depicted in Fig. \ref{fig:exclusionplotnmccoord}.

\begin{figure}[ht]
\centering
\includegraphics[width=\textwidth]{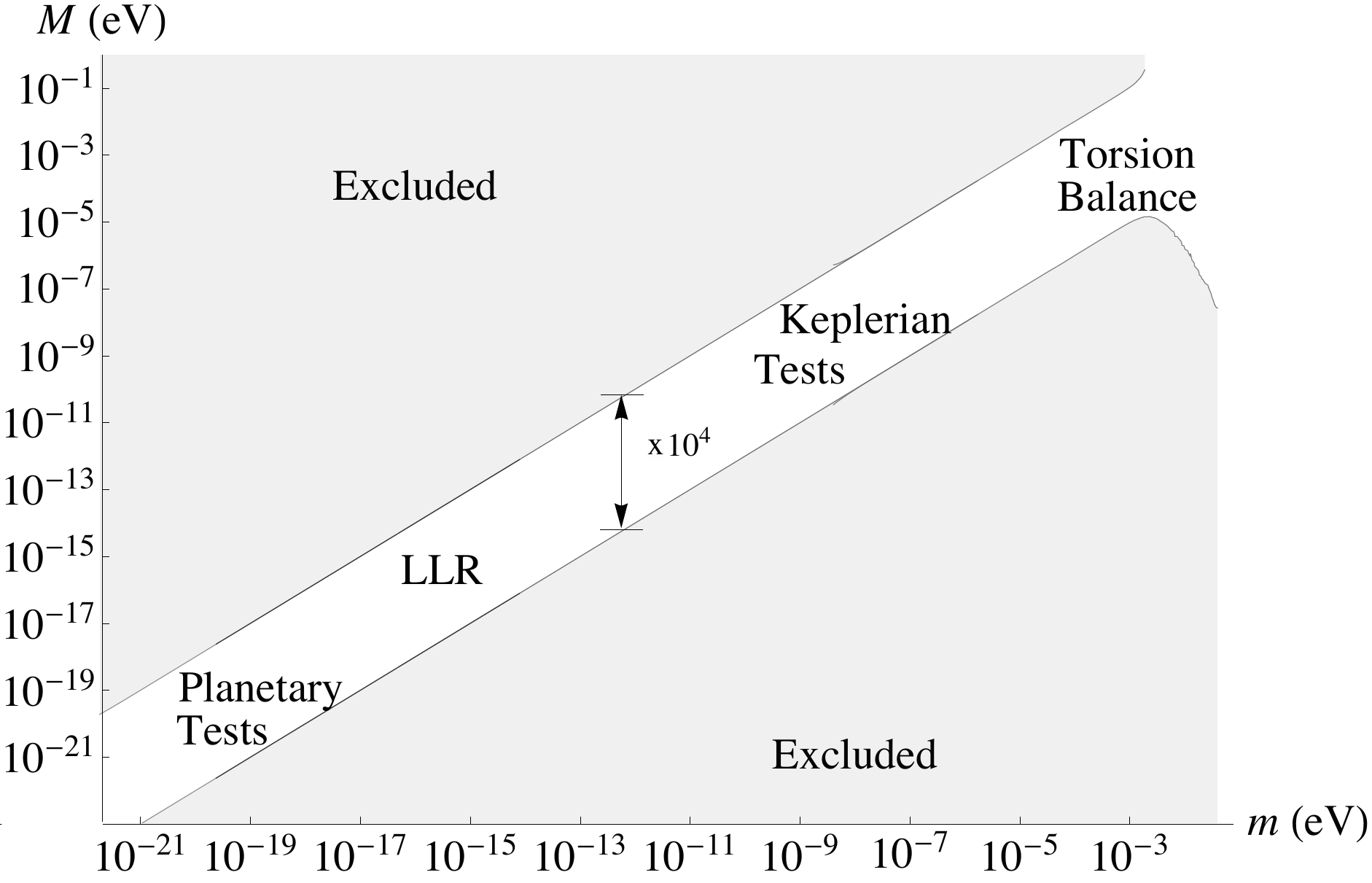}
\caption{Exclusion plot for the characteristic mass scales $M$ and $m$.}
\label{fig:mexclusionplotnmccoord}
\end{figure}

Further insight may be obtained by casting the NMC presented in Eq. \eqref{f(R) equations} as
\begin{equation}\label{transformation, nmc parameter}
f^2(R) = \frac{R}{6 M^2},
\end{equation}
\noindent so that it is characterized by a distinct mass scale $M$, instead of the relative strength parameter $\xi$: by making the transformation $\xi= (m/M)^2/12$, we thus obtain the suggestive form
\begin{equation} \label{transformation, new alpha for new nmc parameter}
\alpha = \dfrac{1}{3} \left[ 1 - \left( \dfrac{m}{M} \right)^2 \right],
\end{equation}
\noindent which, inverting, allows us to plot the exclusion plot for the phase space $(m,M)$ in Fig. \ref{fig:mexclusionplotnmccoord}.


Figs. \ref{fig:exclusionplot}-\ref{fig:mexclusionplotnmccoord} show us that, if $m$ falls within the range $10^{-22} ~{\rm eV} < m < 1~{\rm meV}$ (corresponding to lengthscales $\lambda$ ranging from the millimeter to Solar System scales), then the strong constraints available on the Yukawa strength, $|\alpha| \ll 1$, require that $\xi \sim 1/12$ --- or, equivalently, that both characteristic mass scales are very similar, $m \sim M$.

\subsection{Geodetic precession}

In this section we assume that the Earth can be approximated as a spherically symmetric body.
In order to assess the impact of the obtained expression for $\Phi(r)$, Eq. \eqref{phi solution}, we now consider a gyroscope in a circular orbit with radius $r$ around the Earth.
The intrinsic angular momentum vector $S^\mu = \left(S^0, \vec S \right)$ precesses according to the equation of parallel transport:
\begin{equation}
\frac{dS^\mu}{d\tau} = -\Gamma^\mu_{\nu\sigma}S^\nu\frac{dx^\sigma}{d\tau},
\end{equation}
\noindent where $\tau$ is the proper time. We write the metric tensor around the Earth in rectangular isotropic coordinates, for convenience
\begin{eqnarray} \label{metric in rectangular isotropic coordinates}
ds^2 &=& -\left[ 1 - \frac{2GM_S}{c^2r}\left( 1 + \alpha A(m,R_S)e^{-mr} \right)\right]c^2dt^2 + \nonumber \\
&& \left[ 1 + \frac{2GM_S}{c^2r}\left( 1 -  \alpha A(m,R_S) e^{-mr} \right)\right] dV^2.
\end{eqnarray}
The standard method of computation of gyroscope precession in GR yields for the secular part of
$d\vec S\slash dt$ in NMC gravity, in the slow motion and weak field approximation,
\begin{equation}
\left( \dfrac{d\vec S}{dt} \right)_{{\rm sec}} = \frac{3}{2}\frac{G M_S}{c^2r^3} \left[ 1 - \frac{\alpha A(m,R_S)}{3}(1 + mr)e^{-mr} \right]\left( \vec r \times \vec v \right) \times \vec S,
\end{equation}
\noindent where $\vec r$ is the radius vector of the center of mass of the gyroscope and $\vec v$ is its velocity vector.
By imposing the equality between the acceleration $v^2\slash r$ of the center of mass of the gyroscope
and the sum of the Newton plus Yukawa forces per unit mass, we get
\begin{equation}
vr = \sqrt{G M_S r [ 1 + \alpha A(m,R_S)(1 + mr)e^{-mr} ] }.
\end{equation}
\noindent Since $\left(d\vec S\slash dt\right)_{{\rm sec}} = \vec\Omega_G \times \vec S$, the angular velocity vector $\vec\Omega_G$ of geodetic precession is given by
\begin{eqnarray}
\vec \Omega_G &=& \frac{3}{2}\frac{\left(G M_S \right)^{3\slash 2}}{c^2r^{5\slash 2}} \left[ 1 + \alpha A(m,R_S)(1+mr)e^{-mr} \right]^{1\slash 2} \nonumber\\
&&\times \left[ 1 - \frac{\alpha A(m,R_S)}{3}(1+mr)e^{-mr} \right]\vec n,
\end{eqnarray}
\noindent where $\vec n$ is the unit vector perpendicular to the plane of the orbit.

If $\xi=0$, the above expression reduces to the case for $f(R)$ models, as expected \cite{naf}.

\subsubsection{Gravity Probe B results}

The final results of the Gravity Probe B experiment report an accuracy of 0.28\% in the measurement of geodetic precession \cite{GPB}. This corresponds to the following constraint on NMC gravity parameters:
\begin{equation} \label{GPB constrain, general}
\left\vert \frac{\Omega_G - \Omega_G^{{\rm GR}}}{\Omega_G^{{\rm GR}}} \right\vert < 0.0028,
\end{equation}
\noindent where only the modulus of angular velocity is considered, and $\Omega_G^{{\rm GR}}$ denotes the value of
geodetic precession in GR. Substituting the expression of NMC geodetic precession in this constraint we find
\begin{equation} \label{GPB constrain, NMC}
\Big\vert \sqrt{ 1 + \alpha A(m,R_S) (1+mr)e^{-mr} } \left[ 1 - \frac{\alpha A(m,R_S)}{3}(1+mr)e^{-mr} \right] - 1 \Big\vert  < 0.0028.
\end{equation}
Defining $x \equiv \alpha A(m,R_S) (1 + mr) e^{-mr}$, this is written as
\begin{equation}
\Big\vert \sqrt{ 1 +x } \left( 1 - \frac{x}{3} \right) - 1 \Big\vert  < 0.0028.
\end{equation}
If $x \gg 1$, we get $|x| < 0.04$, which is contradictory. Since substitution shows that $x \sim 1$ breaks the above relation, we are left with the natural constraint $x \ll 1$, so that a first order expansion of the above yields $|x| < 0.0168$. This last condition is translated as
\begin{equation} \label{GPB, alpha for small x}
|\alpha | < \dfrac{0.0168}{1 + mr}\dfrac{e^{mr}}{A(m,R_S)}.
\end{equation}
In order to satisfy the assumption (\ref{conditions for fe}) of continuity of mass density and its derivative across the surface of the Earth, we model the density with a constant value in an interior region ({\it i.e.} mantle plus core) and a sharp transition in a thin crustal layer. When the thickness of the latter tends to zero, the form factor $A(m,R_S) $ converges to the value corresponding to the uniform density model, Eq. (\ref{form factor A(m,R_S) constant limiting}), hence the inequality (\ref{GPB, alpha for small x}) reads
\begin{eqnarray} \label{GPB, alpha for small x limiting}
|\alpha | &<& 0.0168~~,~~mR_\oplus \ll 1 ~~, \\ \nonumber
|\alpha | &<& 0.0112\dfrac{mR_\oplus^2}{ r}e^{m(r-R_\oplus)}~~,~~mR_\oplus \gg 1 ~~,
\end{eqnarray}
\noindent where $R_\oplus \approx 6371$ km is the radius of the Earth. Knowing that the Gravity Probe B orbits the Earth at a height of $\sim 650$ km, this condition can be plotted in the $(\lambda,\alpha)$ exclusion plot, as shown in Fig. \ref{fig:exclusionplot}: we find that it is well-within the already excluded phase space, so that the current bounds on geodetic precession do not add any new constraint on the model parameters.

\subsubsection{Measurement of the LAGEOS II perigee precession}

A recent analysis of the perigee precession of the LAGEOS II satellite reported a much stronger constraint on the strength of the Yukawa perturbation, $\vert\alpha \vert\simeq \vert(1.0\pm 8.9)\vert\times 10^{-12}$, at a range $\lambda = 1\slash m = 6081$ km, very close to one Earth radius \cite{LAGEOSII}: a striking improvement over previous Earth-LAGEOS and Lunar-LAGEOS measurements (at the level of $10^{-5}$ and $10^{-8}$), and comparable to the Lunar Laser Ranging constraint on $\alpha$ for $\lambda \sim 60 R_\oplus$ \cite{LLR}.

Non-gravitational perturbations, mainly
thermal perturbative effects, can strongly affect the precession of the perigee of LAGEOS II: in \cite{LAGEOSII}, solar radiation pressure
and Earth's albedo are taken into account, while Rubincam and Yarkovsky-Schach (YS) thermal effects (which need the satellite spin modeling)
have not been considered. Nevertheless, the residuals in the perigee rate of the satellite are fitted with a linear trend (which represents the secular
total GR precession) plus four periodic terms which correspond to the main spectral lines of the unmodeled YS effect \cite{LAGEOSII}.

This said, if the impressive bound on $\alpha$ quoted above is indeed confirmed, no qualitative changes occur in our previous analysis: as long as the Yukawa coupling strength lies below unity sufficiently, we must have $\xi \sim 1/12 \rightarrow M \sim m$, so that lowering the upper bound on the former only brings the two mass scales of the functions $f^1(R)$ and $f^2(R)$ closer together.

\section{Discussion and Outlook}

In this work we have computed the effect of a NMC model, specified by \eqref{f(R) equations}, in a perturbed weak-field Schwarzschild metric, as depicted in Eq. (\ref{psi solution}) and (\ref{phi solution}). In the weak-field limit, this translates into a Yukawa perturbation to the usual Newtonian potential, with characteristic range and coupling strength
\begin{equation}
\lambda = \dfrac{1}{m}~~~~,~~~~\alpha = \left( \dfrac{1}{3} - 4\xi \right) = \dfrac{1}{3} \left[ 1 - \left( \dfrac{m}{M} \right)^2 \right]~~.
\end{equation}
\noindent This result is quite natural and can be interpreted straightforwardly: a minimally coupled $f(R)$ theory introduces a new massive degree of freedom (as hinted by the equivalence with a scalar-tensor theory \cite{analogy1,analogy2,analogy3}), leading to a Yukawa contribution with characteristic lengthscale $\lambda = 1/m$ and coupling strength $\alpha = 1/3$.

The introduction of a NMC has no dynamical effect in the vacuum, as there is no matter to couple the scalar curvature to: as a result, we do not expect any modification in the range of this Yukawa addition; conversely, a NMC has an impact on the description of the interior of the central body (as illustrated in Refs.\cite{stelobserv,mimlambda,dynimpac2}), leading to a correction to the latter's coupling strength (which has a negative sign since $\LL_m= -\rho$).

Using the available experimental constraints, we find that, for $10^{-22}~{\rm eV} < m < 1~{\rm meV}$ ({\it i.e.} the range $ 10^{-4}~{\rm m} < \lambda < 10^{16}~{\rm m}$), where $|\alpha| \ll 1$, we must have $\xi \sim 1/12$ or, equivalently, that both mass scales $m$ and $M$ of the non-trivial functions $f^1(R)$ and $f^2(R)$ must be extremely close.

If this is the case, the latter relation is not interpreted as an undesirable fine-tuning, but instead is suggestive of a common origin for both non-trivial functions $f^1(R)$ and $f^2(R)$, in line with the argument stating that the model \eqref{model} should arise as a low energy phenomenological approximation to a yet unknown fundamental theory of gravity.

Conversely, for values of $m$ (or $\lambda$) away from the range mentioned above the Yukawa coupling strength $\alpha$ can be much larger than unity, so that $\xi $ can assume any value and the mass scales $m$ and $M$ can differ considerably.

In particular, the Starobinsky inflationary model, which requires the much heavier mass scale $m \approx 3 \times 10^{13}$ GeV $\sim 10^{-6} M_P$, manifests itself at a lengthscale $\lambda \sim 10^{-29}~{\rm m}$. This implies that the generalized preheating scenario posited in Ref. \cite{reheating}, which requires $1<\xi<10^4$, is thus completely allowed by experiment and unconstrained by this work.

By computing the perturbation induced on geodetic precession, we have found that no significant new constraint arises, as this is already included in the existing Yukawa exclusion plot. Furthermore, even considering the much improved precision claimed in a recent study of LAGEOS II --- or, for that matter, any further refinement of $|\alpha| \ll 1$---, no qualitatively new results arise, since this only bridges the gap between $m $ and $M$ ({\it i.e.} narrows the value of $|\xi - 1/12|$).

Finally, a word is due for the so-called chameleon mechanism, first posited in Ref. \cite{chameleon1,chameleon2,chameleon3,chameleon4}, as discussed in Ref. \cite{naf} in relation to $f(R)$ theories. This non-linear effect goes beyond the linear expansion of the modified field equations, and relies on the equivalence between $f(R)$ theories and a scalar-tensor theory with a scalar field $\phi$ proportional to $f_R$, which appears non-minimally coupled to the matter Lagrangian density (in the Einstein frame) \cite{analogy1,analogy2,analogy3,felice}.

As it turns out, the effective potential of this scalar field can be written as $V_{eff}(\phi) = V(\phi) + e^{a\phi}\rho$ (with $a$ an appropriate constant), so that the position of its minimum depends on the density $\rho$, and the mass for the scalar field grows with the density: this is particularly relevant in a cosmological context, where the low background density yields a light, long-ranged field.

Given the above, Ref. \cite{naf} speculates that further computations allowing for this non-linear effect could lead to different constraints on the mass scale $m$ of the adopted quadratic form for $f^1(R)$: quite naturally, the inclusion of a direct coupling between curvature ({\it vis-\`a-vis} the scalar field) and matter only heightens this possibility.

Clearly, this prompts for a future study of the relation between this chameleon mechanism and a NMC model, in the framework of its equivalence with a multi-scalar-tensor theory \cite{multiscalar}.

\section*{Acknowledgments}

The authors thank O. Bertolami for fruitful discussions. The work of J.P. is partially supported by FCT (Funda\c{c}\~ao para a Ci\^encia e a Tecnologia, Portugal) under the project PTDC/FIS/111362/2009. The work of R.M. is partially supported by INFN (Istituto Nazionale di Fisica Nucleare, Italy), as part of the MoonLIGHT-2 experiment in the framework of the research activities of the Commissione Scientifica Nazionale n. 2 (CSN2).

\bibliographystyle{elsarticle-num}
\bibliography{nmcperturbation}

\end{document}